# Exact decomposition of non-Markovian dynamics in open quantum systems

Mariia Ivanchenko[a,#], Peter L. Walters[a,#], Fei Wang[a,b*]

[a]Department of Chemistry and Biochemistry, George Mason University, 4400 University Drive, Fairfax, Virginia 22030, USA

[b]Quantum Science and Engineering Center, George Mason University, 4400 University Drive, Fairfax, Virginia 22030, USA

*Email: fwang22@gmu.edu

#: These authors contribute equally to the work.

## Abstract

In this work, we developed a rigorous procedure for mapping the exact non-Markovian propagator to the generalized Lindblad form. It allows us to exact the negative decay rate that is the indicator of the non-Markovian effect. As a consequence, we can investigate the influence of the non-Markovian bath on the system's properties such as coherence and equilibrium state distribution. The understanding of the non-Markovian contribution to the dynamical process points to the possibility of leveraging non-Markovianity for quantum control.

## I.   Introduction

The quantum dynamics of open quantum systems are the underpinnings of the charge and excitation energy transfer processes in the condensed phase environment.[1–8] The dynamical behavior can be Markovian or non-Markovian. Non-Markovianity refers to the fact that the dynamics of the past influences its trajectory of the present and the future. This memory effect makes the simulation and the analysis of non-Markovian quantum dynamics a more demanding task compared to its Markovian counterpart. Intense efforts have been made in developing numerically exact methods for simulating non-Markovian quantum dynamics.[9–18] On the other hand, non-Markovian measures and witnesses have been proposed to quantify and detect non-Markovian quantum processes.[19] What is still lacking is a rigorous procedure that partitions the dynamics into a Markovian and a non-Markovian component, and the subsequent analysis to pinpoint the unique contribution of the non-Markovian part to the quantum transport properties. With these needs in mind, we have devised a systematic approach for separating the non-Markovian contribution from the dynamical equation and provided a framework for post-calculation analysis of the non-Markovian effect on the dynamics. In the following, we will focus on the Feynman-Vernon's influence functional formulation and the generalized Lindblad equation.

## II.   From the exact propagator to the Lindblad form
## A.   Linear map

The reduced density matrix (RDM) $\rho_s$ at time $t$ can be obtained from the time evolution of the initial state,

$$\rho_s(t) = \phi_t[\rho_s(0)] \qquad (1)$$

in which $\phi_t$ is a linear map or superoperator that is complete positive and trace-preserving (CPTP).[20] Several numerically exact methods are available for constructing such a superoperator, such as QuAPI,[9,10] QCPI,[11,12] SMatPI,[21,22] HEOM,[13] ML-MCTDH,[14,15] TNPI,[16,17] etc. In this work, we use TNPI[16] to generate the linear map.

## B. Path integral formulation

We resort to a multi-state system linearly coupled to its harmonic bath as our model system in the following discussion, as it has been shown to be widely applicable for simulating non-Markovian quantum dynamics in many condensed phase systems.[3-8] In the Feynman's path integral formulation, the time evolution of the RDM has the form

$$\rho_s(s_t^+, s_t^-) = \int \mathcal{D}s^+ \int \mathcal{D}s^- \exp\left\{\frac{i}{\hbar}(S[s^+] - S[s^-])\right\} IF[s^+, s^-]\langle s_0^+|\rho_0(0)|s_0^-\rangle \quad (2)$$

where

$$IF[s^+, s^-] = \exp\left\{-\frac{1}{\hbar}\int_0^t dt' \int_0^{t'} dt'' \, [s^+(t') - s^-(t')][\alpha(t' - t'')s^+(t'') - \alpha^*(t' - t'')s^-(t'')]\right\} \quad (3)$$

is the Feynman-Vernon's influence functional,[2,23] with the $s^+$ and $s^-$ the forward and backward paths, respectively. $S[s^+]$ and $S[s^-]$ are the action integrals of the free system, and $\langle s_0^+|\rho_0(0)|s_0^-\rangle$ is the initial state. The free propagator multiplied by the influence functional defines the superoperator. The non-local memory kernel $\alpha(t' - t'')$ is the root of non-Markovianity, and can be obtained from the bath response function[2]

$$\alpha(t) = \frac{1}{\pi}\int_0^\infty d\omega J(\omega)\left[\coth\left(\frac{\hbar\omega\beta}{2}\right)\cos(\omega t) - i\sin(\omega t)\right] \quad (4)$$

in which the spectral density $J(\omega)$ is defined as[24]

$$J(\omega) = \frac{\pi}{2}\sum_j \frac{c_j^2}{m_j \omega_j}\delta(\omega - \omega_j) \quad (5)$$

Here, $\omega_j$ is the bath frequency, and $c_j$ is the coupling strength between the system and bath.

## C. Converting the propagator from the path integral expression to the Lindblad form[25,26]

The goal is that given $\phi_t$ in equation (1), formulate it in the following form,

$$\dot\rho(t) = \Lambda_t[\rho(t)] := -\frac{i}{\hbar}[H(t), \rho(t)] + \sum_k \gamma_k(t)\left(L_k(t)\rho(t)L_k^\dagger(t) - \frac{1}{2}\{L_k^\dagger(t)L_k(t), \rho(t)\}\right) \quad (6)$$

where $\{\}$ is the anti-commutator. Unlike the conventional Lindblad equation,[1] the decay rate $\gamma_k(t)$ in equation (6) can have a negative value which indicates non-Markovianity.

We decompose the task into three steps. First, for a Hilbert space of dimension $d$, we express $\phi_t$ in the orthonormal basis of Hermitian operators $\{G_m\}$ such that

$$G_0 = \frac{\mathbb{I}}{\sqrt{d}}; \qquad G_m = G_m^\dagger; \qquad \text{Tr}[G_m G_n] = \delta_{mn} \quad (7)$$

in which $\mathbb{I}$ is the identity operator. There are total of $N = d^2$ such basis operators.

To relate a linear map $\phi_t$ to the corresponding $\Lambda_t$, we define

$$F_{kl}(t) := \text{Tr}[G_k \, \phi_t(G_l)] \tag{8}$$

and then solve

$$B(t) = \dot{F}(t) F^{-1}(t) \tag{9}$$

This equation provides a relationship between the linear maps $\phi_t$ and $\Lambda_t$, where each element of the matrix $B(t)$ is

$$B_{kl}(t) = \text{Tr}[G_k \, \Lambda_t(G_l)] \tag{10}$$

Secondly, we find the operator-sum representation of the $\Lambda_t$ by computing

$$D_{ij}(t) = \sum_{k,l=0}^{N-1} B_{kl}(t) \, \text{Tr}\left[G_l G_i G_k G_j\right] \tag{11}$$

Then

$$\Lambda_t(\rho) = \sum_{k,l=0}^{N-1} D_{ij}(t) G_i \rho G_j \tag{12}$$

It is easy to show that $D_{ij}(t)$ is Hermitian.

Thirdly, we express $\Lambda_t(\rho)$ in a canonical Lindblad form. We may define

$$A(t) := \frac{1}{2} \frac{D_{00}(t)}{d} + \sum_{i=1}^{N-1} \frac{D_{i0}(t)}{\sqrt{d}} G_i \tag{13}$$

Then the Hamiltonian can be identified as

$$H(t) := \frac{i}{2} \hbar \left( A(t) - A^\dagger(t) \right) \tag{14}$$

Subsequently, the dynamical equation can be expressed as

$$\dot{\rho}(t) = -\frac{i}{\hbar}[H(t), \rho(t)] + \sum_{i,j=1}^{N-1} D_{ij}(t) \left( G_i \rho G_j - \frac{1}{2}\{G_j G_i, \rho\} \right) \tag{15}$$

where $D_{ij}(t)$ $(i,j \neq 0)$ is the decoherence matrix.

As $D_{ij}(t)$ $(i,j \neq 0)$ is Hermitian, it can be diagonalized,

$$D_{ij}(t) = \sum_{k=1}^{N-1} U_{ik}(t) \gamma_k(t) U^*_{jk}(t) \tag{16}$$

where $\gamma_k(t)$ are the eigenvalues of $D_{ij}(t)$ $(i,j \neq 0)$.

The Lindblad operators are defined as

$$L_k(t) := \sum_{i=1}^{N-1} U_{ik}(t)\, G_i \qquad (17)$$

which recovers the form of equation (6).

The Lindlad form thus constructed we call the *generalized* Lindblad equation.

## III. Characterization of non-Markovianity
### A. Non-Markovian measures

Although many theoretical methods have been proposed to detect non-Markovianity, most of them are classified as non-Markovian witnesses.[27–34] A witness may detect non-Markovianity only in certain cases whereas a non-Markovian measure[25,34–37] is a quantity that is non-zero for all non-Markovian cases. Some measures such as those based on optimization[35,36] are difficult to compute in practice, whereas the ones based on the violation of the complete positivity[34] and on the negative decay rate[25] are amenable to calculations within the framework of the Lindblad equation.

**(a) RHP measure**

We first introduce the RHP (Rivas, Huegla and Plenio) measure[34] which is based the quantification of how much the dynamics deviates from complete positivity (CP).

The short-time propagator for Markovian dynamics is a complete positive and trace-preserving (CPTP) map.[38] Equivalently, the Choi matrix[39,40] of the Markovian short-time propagator is positive semidefinite and the trace-norm of such Choi matrix is exactly one; on the contrary, the trace-norm of the Choi matrix of the non-Markovian short-time propagator is larger than one. [19,34] The precise mathematical statement can be formulated with the aid of defining a maximally entangled state as

$$|\Phi\rangle := \frac{1}{\sqrt{d}} \sum_{n=0}^{d-1} |n\rangle|n\rangle \qquad (18)$$

Since the short-time propagator can be expressed from the Lindblad superoperator as $\mathbb{I} + \varepsilon \Lambda_t$, with $\varepsilon$ being an arbitrarily small value, then Choi matrix associated with this short-time propagator is

$$[\mathbb{I} + \varepsilon(\Lambda_t \otimes \mathbb{I})]|\Phi\rangle\langle\Phi| \qquad (19)$$

The trace-norm condition indicates that

$$\|[\mathbb{I} + \varepsilon(\Lambda_t \otimes \mathbb{I})]|\Phi\rangle\langle\Phi|\|_1 \begin{cases} = 1 & \text{iff Markovian} \\ > 1 & \text{iff non} - \text{Markovian} \end{cases} \qquad (20)$$

where $\|\cdot\|_1$ is the trace norm or the Schatten 1-norm defined as

$$\|A\|_1 := \operatorname{Tr}\sqrt{AA^\dagger} \qquad (21)$$

Equation (21) is also equal to the sum of the singular values of $A$.

The RHP measure[34] quantify the non-Markovianity by defining a function $g(t)$ such that

$$g(t) := \lim_{\varepsilon \to 0^+} \frac{\|[\mathbb{I} + \varepsilon(\Lambda_t \otimes \mathbb{I})]|\Phi\rangle\langle\Phi|\|_1 - 1}{\varepsilon} \tag{22}$$

**(b) Decay rate measure**

It has been shown that the negative value of the decay rate $\gamma_k(t)$ can also serve as a true measure of non-Markovianity.[25] Specifically, the non-Markovianity of a specific Lindblad channel can be quantified by

$$f_k(t) := \max[0, -\gamma_k(t)] \geq 0 \tag{23}$$

whereas the non-Markovianity for the dynamics at time $t$ can be calculate by

$$f(t) := \sum_{k=1}^{N-1} f_k(t) \tag{24}$$

The decay rate measure provides a way to find the *conjugate* Markovian dynamics from the non-Markovian one by eliminating the negative values of the decay rates in the generalized Lindblad equation.

The equivalence between the decay rate measure and the RHP measure has been elegantly proved, stating that[25]

$$f(t) = \frac{d}{2} g(t) \tag{25}$$

where $d$ is the dimension of the state space. The advantage of using the Lindblad form is that one can define a "non-Markov index" as the number of negative decay rate,[25] quantifying the dimension of space permeated by non-Markovian dynamics. In addition, it has been demonstrated that the minimal rate of isotropic noise that must be added to produce Markovian dynamics is given by the most negative decay rate.[25,35]

**B. Bloch sphere representation**

The Bloch sphere can conveniently describe the qubit state and its dynamical process. The density matrix of a qubit can be expressed as

$$\rho = \frac{(\mathbb{I} + \vec{r} \cdot \vec{\sigma})}{2} \tag{26}$$

where $\vec{r}$ is a real three-component vector and $\vec{\sigma}$ are the Pauli matrices. Any trace-preserving quantum operation on a qubit corresponds to an affine transformation,

$$\vec{r}' = M\vec{r} + \vec{C} \tag{27}$$

where $\vec{C}$ is a constant vector that accounts for translation of the Bloch sphere and $M$ is a $3 \times 3$ real matrix. The polar decomposition of the matrix $M$ is given by

$$M = OS \tag{28}$$

where $O$ is a real orthogonal matrix with determinant 1 and $S$ is a real symmetric matrix.[20] Therefore, the linear transformation $M$ represents first a stretching and/or compression of the Bloch sphere along the principal axes (defined by $S$), and then a rotation (determined by $O$). We can clearly visualize the difference between Markovian and non-Markovian process for qubit dynamics and assess the non-Markovian contribution to the linear transformation and to the translation.

The Bloch volume evolves according to $V(t) = V(0) \det M(t)$.[31] Hence, the increase of the Bloch volume is a witness to the non-Markovianity. It can be shown that

$$\frac{d}{dt} \det M(t) = \text{Tr}[B(t)] \det M(t) \tag{29}$$

where $B(t)$ is given in equation (10). The Bloch volume increase implies $\text{Tr}[B(t)] > 0$. Moreover,[25]

$$\text{Tr}[B(t)] = -d \sum_{k=1}^{N-1} \gamma_k(t) \tag{30}$$

pointing to the benefit of working with the Lindblad form.

The translational vector $\vec{C}$ can be accessed by

$$C_i = \text{Tr}[G_i\, \phi_t(\mathbb{I})]/d, \quad i > 0 \tag{31}$$

in which $\phi_t$ could be the propagator for the non-Markovian dynamics or its conjugate Markovian one.

## IV. Results and Discussions

In what follows, we present the simulation results using the spin-boson model, with the system-bath Hamiltonian given by

$$H = \hbar\Omega\sigma_x + \hbar\varepsilon\sigma_z + \sum_j \left[\frac{p_j^2}{2m_j} + \frac{1}{2}m_j\omega_j^2\left(x_j - \frac{c_j\sigma_z}{m_j\omega_j^2}\right)^2\right] \tag{32}$$

where $p_j$ and $x_j$ denote the bath's momentum and position, respectively, and $c_j$ denotes the system-bath coupling strength. We choose the spectral density to have the Ohmic form,

$$J(\omega) = \frac{\pi}{2}\hbar\xi\omega e^{-\omega/\omega_c} \tag{33}$$

which gives a continuous version of equation (5). The dimensionless $\xi$ is the Kondo parameter that determines the strength of the system-bath coupling, and $\omega_c$ is the cutoff frequency of the bath. Our simulations and analysis explore three parameter regimes. Figure 1-6 has the parameters of $\Omega = 1, \beta = 5, \xi = 0.1, \omega_c = 7.5$, and $\varepsilon = 0$, which shows damped oscillation of a symmetric two-state system. Figure 7-12 has the parameters of $\Omega = 1, \beta = 5, \xi = 0.1, \omega_c = 7.5$, and $\varepsilon = 1$, which shows damped oscillation of an asymmetric two-state system. Figure 13-18 has the parameters of $\Omega = 1, \beta = 5, \xi = 0.5, \omega_c = 7.5$, and $\varepsilon = 1$, which shows dissipative decay of an asymmetric two-state system. In each parameter set, it sequentially displaces six items: (1) the matching results of the generalized Lindblad equation with the exact benchmark, (2) the decay rates, (3) the separation of the conjugate Markovian dynamics from its corresponding non-Markovian one, (4) comparison of the Bloch sphere dynamics between non-Markovian and its conjugate Markovian process, (5) difference in the translational vector between non-Markovian and its conjugate Markovian dynamics, and (6) the matching of the decay rate measure and the RHP measure.

Figures 1, 7 and 13 clearly demonstrate that the Lindblad form constructed by the procedures outlined in Section II C. can produce exactly the correct dynamical behavior, confirming the reliability of using the canonical Lindblad formalism to simulate non-Markovian quantum dynamics. The exact benchmark simulations are obtained from the TNPI[16] method. Figures 2, 8 and 14 show the decay rates $\gamma_k(t)$ in the Lindblad equation. They are time-dependent, and the negative values indicate non-Markovianity. The perfect matching of the decay rate measure with the RHP measure is shown in Figure 6, 12 and 18. It clearly points to the robustness of using the Lindblad form to analyze non-Markovian quantum dynamics. In what follows, we will give a thorough analysis of the results for parameter set 1, and briefly point to the similarities and differences for parameter sets 2 and 3.

Figure 3 shows the separation of Markovian dynamics from the fully non-Markovian one by setting the negative values of the decay rate to zero. The plots in A, B and C are the population dynamics, the real part of the off-diagonal element of the RDM, and the imaginary part of the off-diagonal of the RDM, respectively. Two distinctive features emerge from the comparison. First, the non-Markovian bath preserves the coherence better than the Markovian bath, as can be seen in Figure 1 with the larger amplitude of the oscillation in the non-Markovian case. In other words, non-Markovianity has the benefit of hedging against decoherence and dissipation. This can also be seen from Figure B and C, the coherence part of the RDM. The imaginary part of the off-diagonal element of the RDM is related to the time derivative of the population,[41] and therefore Figure C correctly correlates with the behavior in Figure A. The second feature arises from the observation of Figure B. The real part of the off-diagonal element of the RDM is equal to one-half of the population difference between the two eigenstates.[42] The non-Markovian bath has the effect of opening up the energy gap, and therefore leading to the possibility of tuning the system's energy levels by leveraging non-Markovianity. Since the real part of the off-diagonal stabilizes at a larger value in the non-Markovian case, it has the effect of coherence trapping.[43]

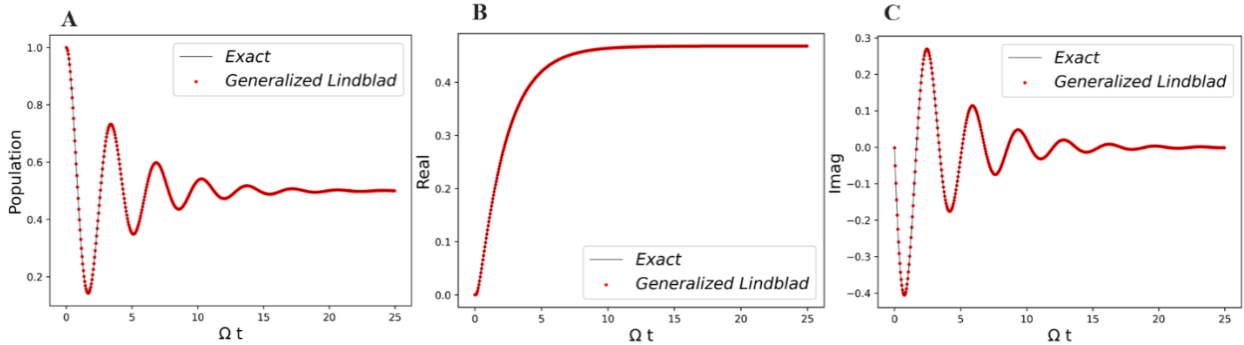

**Figure 1**. Generalized Lindblad equation result compared with the exact simulation. **A**. The population dynamics of the symmetric two-state system. **B.** The real part of the off-diagonal element of the RDM. **C.** The imaginary part of the off-diagonal element of RDM. Parameters $\Omega = 1, \beta = 5, \xi = 0.1, \omega_c = 7.5, \varepsilon = 0$.

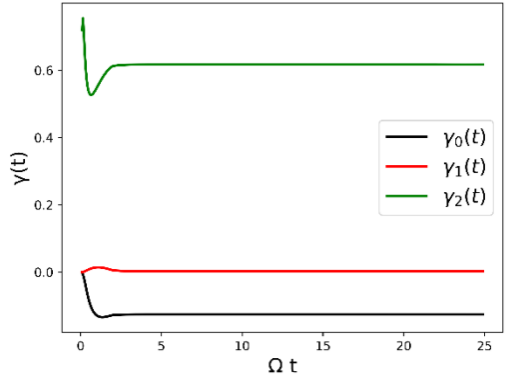

**Figure 2**. Time-dependent γ coefficients (decay rates). Parameters $\Omega = 1, \beta = 5, \xi = 0.1, \omega_c = 7.5, \varepsilon = 0$.

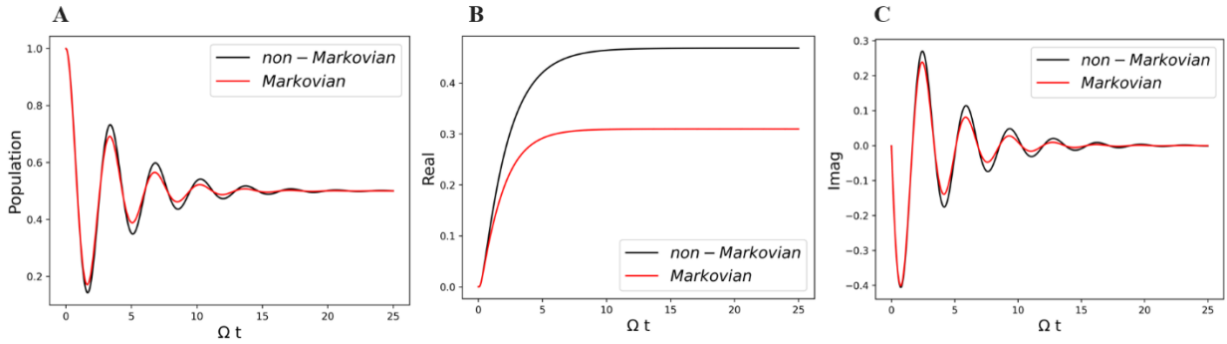

**Figure 3**. The comparison of the non-Markovian (black curve) and Markovian (red curve) dynamics. **A.** Population dynamics of the symmetric two-state system. **B.** The real part of the off-diagonal element of the RDM. **C.** The imaginary part of the off-diagonal element of RDM. Parameters $\Omega = 1, \beta = 5, \xi = 0.1, \omega_c = 7.5, \varepsilon = 0$.

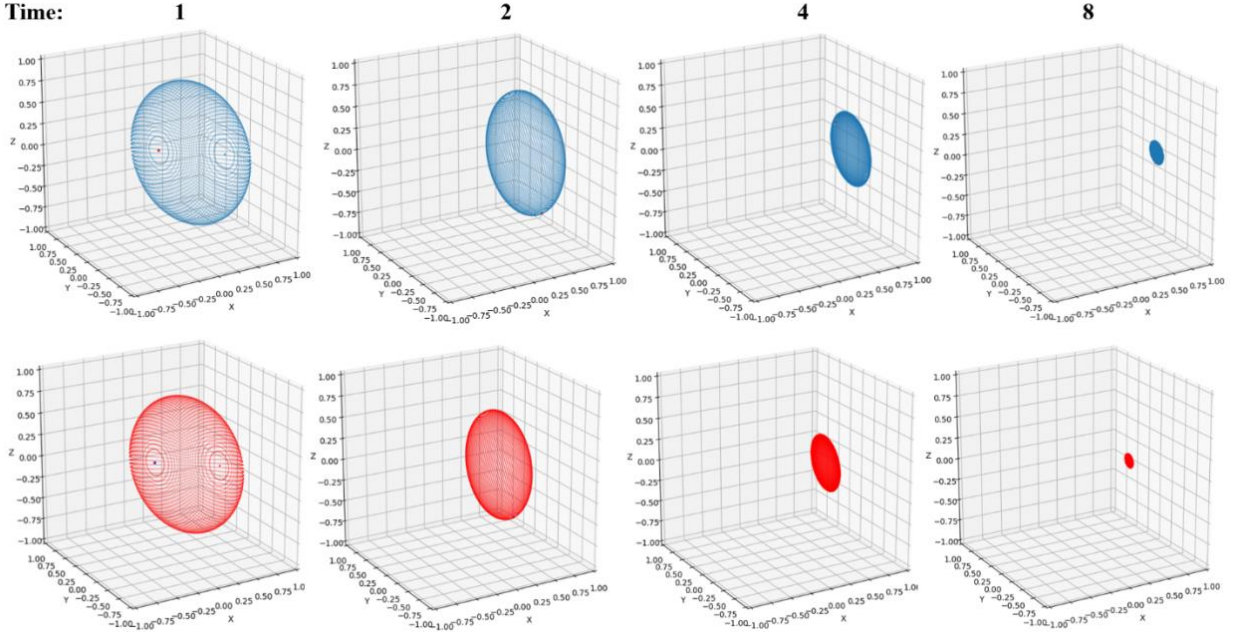

**Figure 4**. The Bloch sphere under non-Markovian (blue) and Markovian (red) evolution at $t = 1$, $t = 2$, $t = 4$, and $t = 8$. Parameters $\Omega = 1, \beta = 5, \xi = 0.1, \omega_c = 7.5, \varepsilon = 0$.

Figure 4 gives a pictorial representation of the non-Markovian and Markovian time evolution by plotting the Bloch sphere change over time. As discussed in Section III. B, the rate of change is determined by the $M$ matrix in equation (27). It can be clearly observed that the Markovian dynamics suffers more from the dissipation and decoherence from the bath, with the Bloch volume shrinking at a faster rate. Both spheres will eventually collapse to a point that has zero value in the y and z component, and finite value in the x component. This exactly matches the behaviors in Figure 3. The population reaches the 50-50 equilibrium state, indicating the z component goes to zero. The imaginary part of the off-diagonal element approaches zero, indicating the y component is zero. Only the real part of the off-diagonal element reaches a finite value, which contributes to the $\sigma_x$ component in equation (26).

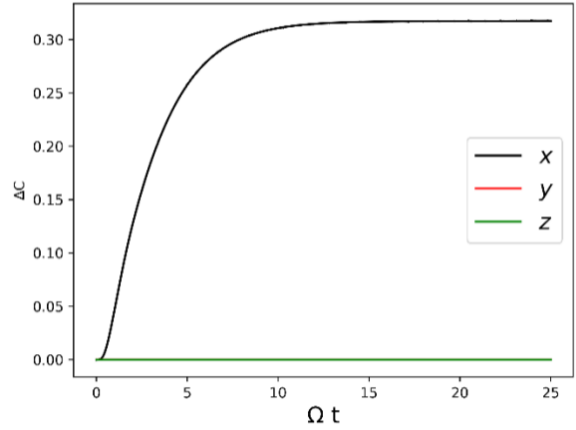

**Figure 5**. Non-Markovian and Markovian comparison of the displacement of the center of the Bloch sphere over time. Parameters $\Omega = 1, \beta = 5, \xi = 0.1, \omega_c = 7.5, \varepsilon = 0$.

Figure 5 gives the difference in the translational vector $\vec{C}$ in equation (27) between the non-Markovian and its conjugate Markovian dynamics. The difference lies only in the x component in this parameter regime. This feature exactly matches the behavior in Figure 3B. It is interesting to compare this non-Markovian feature with the witness defined by the Bloch volume increase (equation 29 and 30). We can see from Figure 2 that the sum of these decay rates gives a positive value, hence the Bloch volume keeps decreasing over time. It points to an interesting fact that the non-Markovianity in all the parameters regimes we chose is solely encoded in the translational vector, not the Bloch volume. This is also true for Figure 8 and 14 accidentally. It also raises caution about the use of the Bloch volume increase as a witness to non-Markovianity.

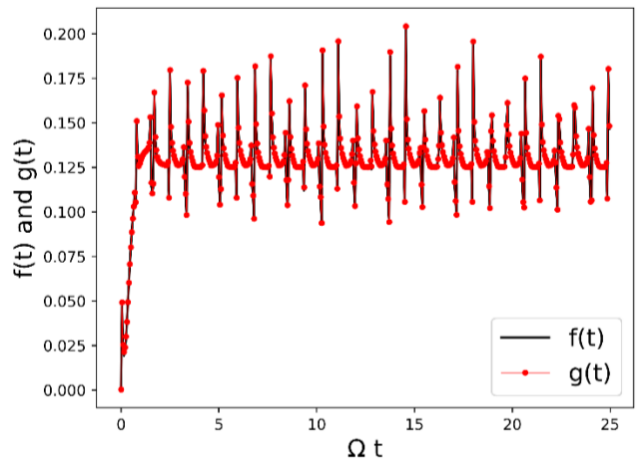

**Figure 6**. Comparison of the decay rate measure $f(t)$ with the RHP measure $g(t)$ for the two-state system. Parameters $\Omega = 1, \beta = 5, \xi = 0.1, \omega_c = 7.5, \varepsilon = 0$.

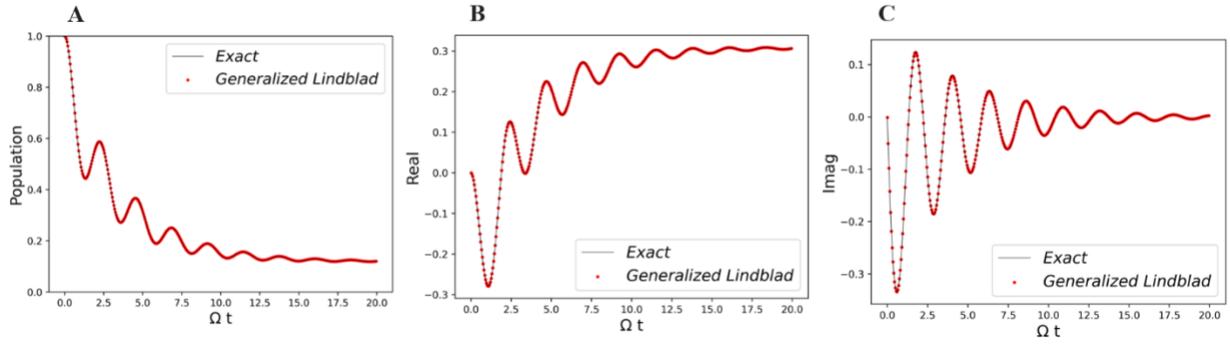

**Figure 7.** Generalized Lindblad equation result compared with the exact simulation. **A.** The population dynamics of the symmetric two-state system. **B.** The real part of the off-diagonal element of the RDM. **C.** The imaginary part of the off-diagonal element of RDM. Parameters $\Omega = 1, \beta = 5, \xi = 0.1, \omega_c = 7.5, \varepsilon = 1$.

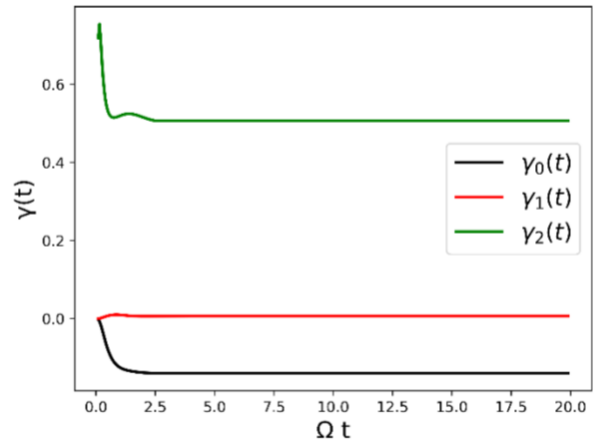

**Figure 8.** Time-dependence of the $\gamma$ coefficients (decay rates). Parameters $\Omega = 1, \beta = 5, \xi = 0.1, \omega_c = 7.5, \varepsilon = 1$.

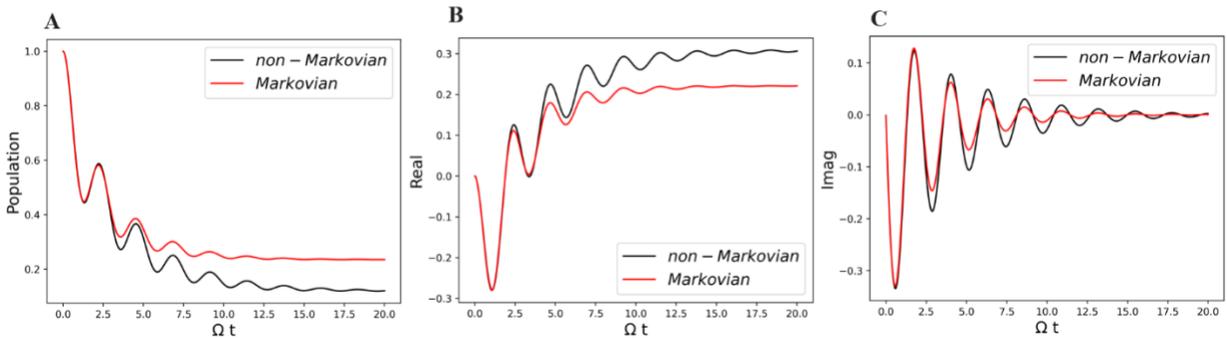

**Figure 9.** The comparison of the non-Markovian (black curve) and Markovian (red curve) dynamics. **A.** Population dynamics of the asymmetric two-state system. **B.** The real part of the off-diagonal element of the RDM. **C.** The imaginary part of the off-diagonal element of RDM. Parameters $\Omega = 1, \beta = 5, \xi = 0.1, \omega_c = 7.5, \varepsilon = 1$.

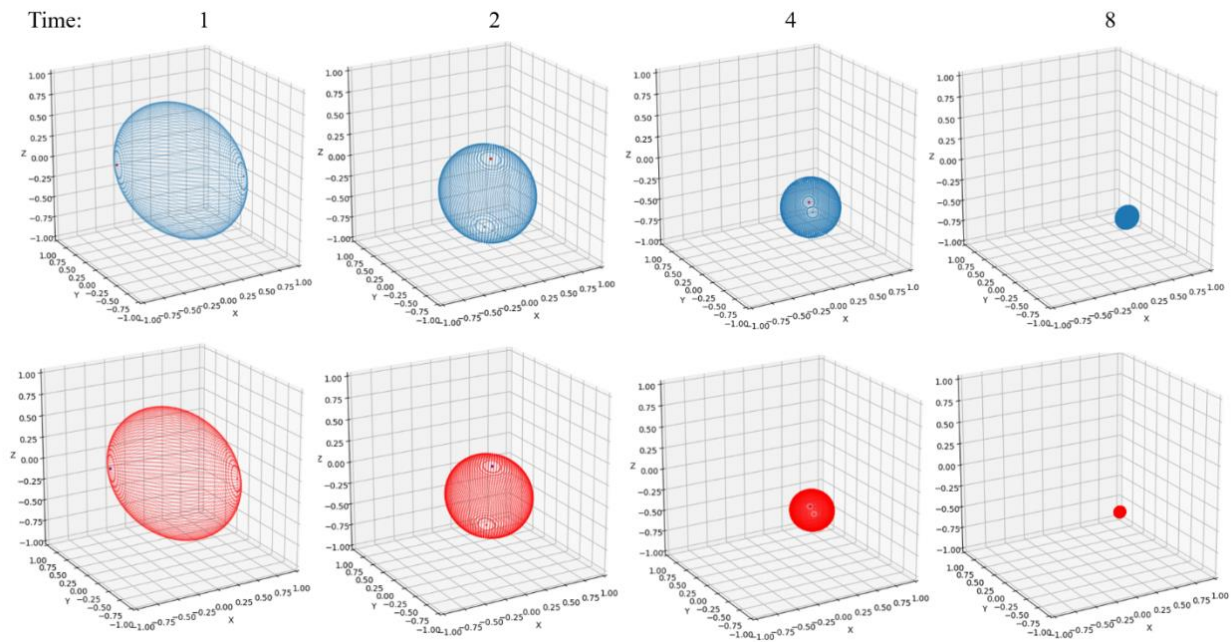

**Figure 10**. The Bloch sphere under non-Markovian (blue) and Markovian (red) evolution at $t = 1$, $t = 2$, $t = 4$, and $t = 8$. Parameters $\Omega = 1, \beta = 5, \xi = 0.1, \omega_c = 7.5, \varepsilon = 1$.

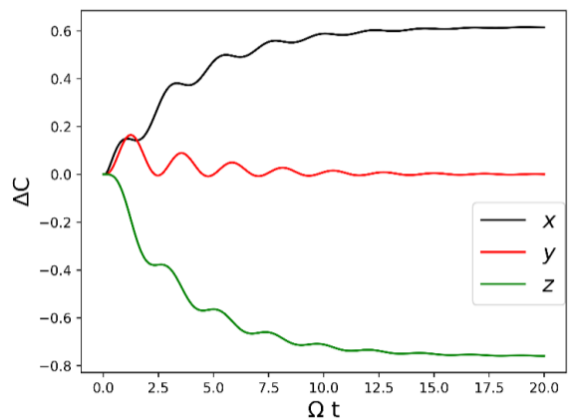

**Figure 11**. Non-Markovian and Markovian comparison of the displacement of the center of the Bloch sphere over time. Parameters $\Omega = 1, \beta = 5, \xi = 0.1, \omega_c = 7.5, \varepsilon = 1$.

Figure 9 demonstrates the difference between non-Markovian and its conjugate Markovian dynamics for an asymmetric two-state system. Besides the coherence preservation and coherence trapping features observed for the symmetric system, Figure 9A clearly indicates that the non-Markovian bath alters the equilibrium constant. Particularly in this case, it drives the reaction to a higher percentage yield. Figure 15A shows the same trend with a different parameter set. Therefore, another usefulness of the non-Markovian bath arises from its ability to shift chemical equilibrium. Figure 11 restates the properties of equilibrium shift (z component) and the coherence trapping (x component).

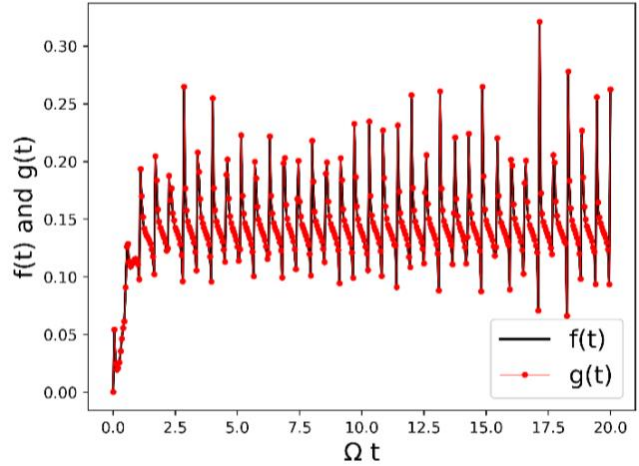

**Figure 12**. Comparison of the decay rate measure $f(t)$ with the RHP measure $g(t)$ for the two-state system. Parameters $\Omega = 1, \beta = 5, \xi = 0.1, \omega_c = 7.5, \varepsilon = 1$.

Whereas Figure 1-12 show damped oscillations, Figure 13-18 show pure dissipative decay. They reveal the same features thus discussed.

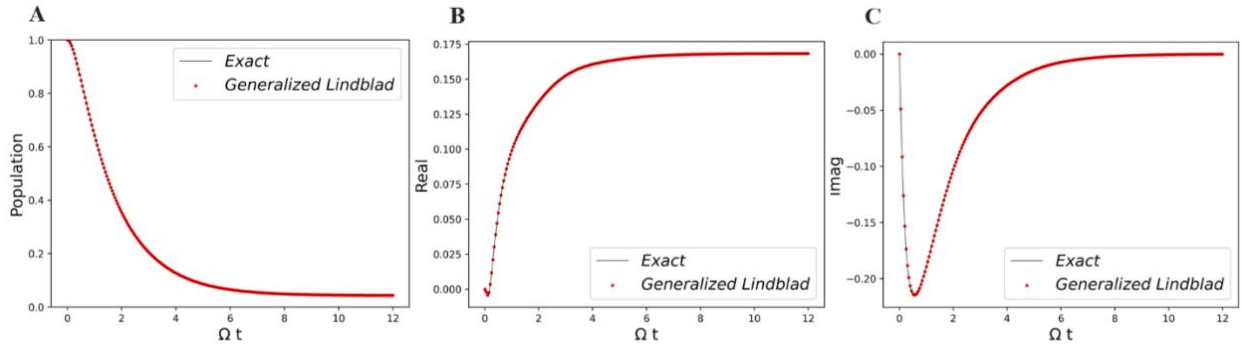

**Figure 13**. Generalized Lindblad equation result compared with the exact simulation. **A.** The population dynamics of the symmetric two-state system. **B.** The real part of the off-diagonal element of the RDM. **C.** The imaginary part of the off-diagonal element of RDM. Parameters $\Omega = 1, \beta = 5, \xi = 0.5, \omega_c = 7.5, \varepsilon = 1$.

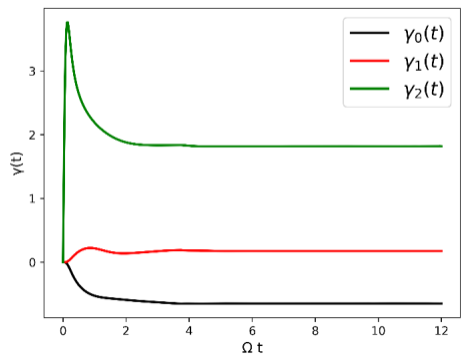

**Figure 14**. Time-dependence of the $\gamma$ coefficients (decay rates). Parameters $\Omega = 1, \beta = 5, \xi = 0.5, \omega_c = 7.5, \varepsilon = 1$.

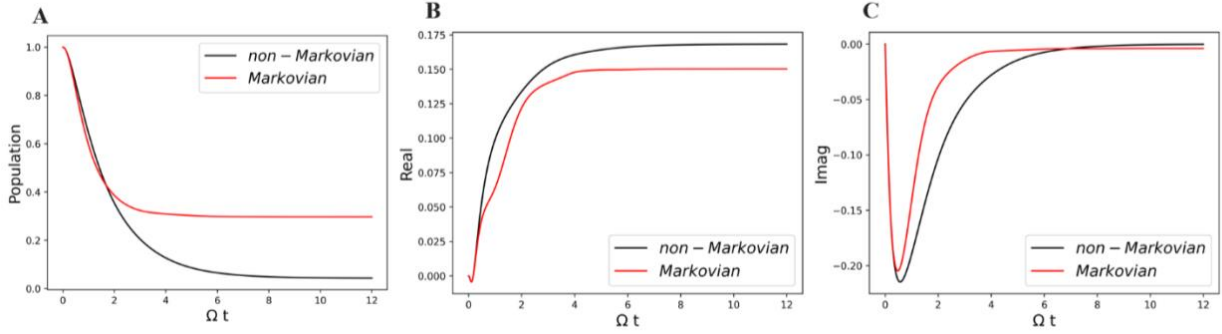

**Figure 15**. The comparison of the non-Markovian (black curve) and Markovian (red curve) dynamics. **A**. Population dynamics of the symmetric two-state system. **B**. The real part of the off-diagonal element of the RDM. **C**. The imaginary part of the off-diagonal element of RDM. Parameters $\Omega = 1, \beta = 5, \xi = 0.5, \omega_c = 7.5, \varepsilon = 1$.

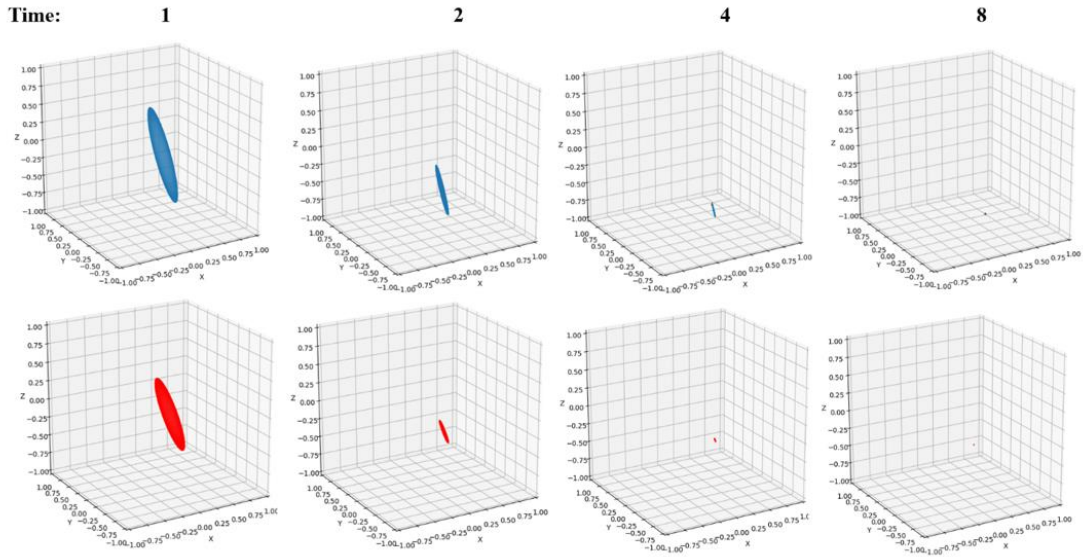

**Figure 16**. The Bloch sphere under non-Markovian (blue) and Markovian (red) evolution at $t = 1$, $t = 2$, $t = 4$, and $t = 8$. Parameters $\Omega = 1, \beta = 5, \xi = 0.5, \omega_c = 7.5, \varepsilon = 1$.

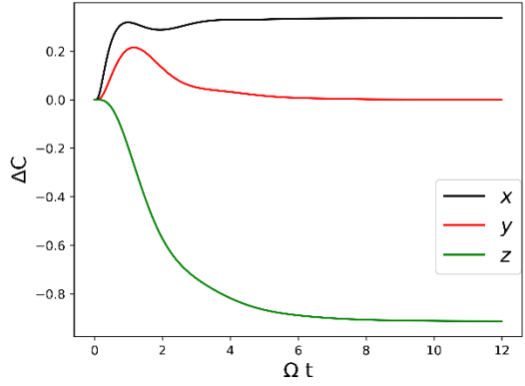

**Figure 17**. Non-Markovian and Markovian comparison of the displacement of the center of the Bloch sphere over time. Parameters $\Omega = 1, \beta = 5, \xi = 0.5, \omega_c = 7.5, \varepsilon = 1$.

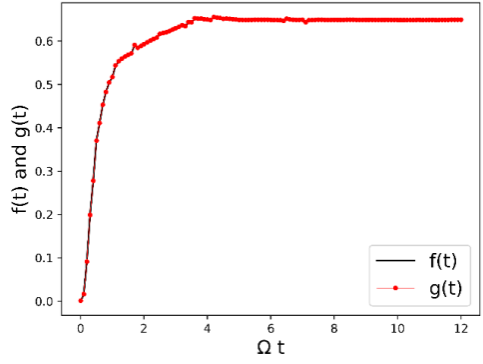

**Figure 18**. Comparison of the decay rate measure $f(t)$ with the RHP measure $g(t)$ for the two-state system. Parameters $\Omega = 1, \beta = 5, \xi = 0.5, \omega_c = 7.5, \varepsilon = 1$.

## V. Conclusion

In this work, we have presented a systematic approach to convert the exact non-Markovian dynamical map to the Lindblad form. The immediate structure of the Lindblad dissipator allows to extract the decay rate and quantify non-Markovianity. The separation of the negative decay rates gives us the ability to investigate the effect of the non-Markovian bath on the properties and the dynamics of the system. We have shown that the non-Markovian bath can better protect coherence, change the energy gap and shift the equilibrium. It is not hard to infer that other quantum properties such as the quantum transport rate and entanglement can be equally examined in the same way. We have also provided a pictorial representation of the qubit dynamics by the Bloch sphere time-evolution and the associated affine transformation. This visualization gives a more direct impression of the dynamical process influenced by the non-Markovian bath. We believe that the approach we presented is useful and practical for thoroughly analyzing non-Markovian quantum dynamics in open quantum systems. In addition, non-Markovianity can be leveraged as a resource[30,44] to achieve quantum control through reservoir engineering.[45–49]

## Acknowledgements

This work is supported by the National Science Foundation (NSF) under the Award 2320328, and the George Mason University's startup fund.